# Gradually Atom Pruning for Sparse Reconstruction and Extension to Correlated Sparsity

Seyed Hossein Hosseini and Mahrokh G. Shayesteh
Department of Electrical Engineering, Urmia University, Urmia, Iran
Emails: st_h.hosseini@urmia.ac.ir ; m.shayesteh@urmia.ac.ir

*Abstract*— We propose a new algorithm for recovery of sparse signals from their compressively sensed samples. The proposed algorithm benefits from the strategy of gradual movement to estimate the positions of non-zero samples of sparse signal. We decompose each sample of signal into two variables, namely "value" and "detector", by a weighted exponential function. We update these new variables using gradient descent method. Like the traditional compressed sensing algorithms, the first variable is used to solve the Least Absolute Shrinkage and Selection Operator (Lasso) problem. As a new strategy, the second variable participates in the regularization term of the Lasso ($l_1$ norm) that gradually detects the non-zero elements. The presence of the second variable enables us to extend the corresponding vector of the first variable to matrix form. This makes possible use of the correlation matrix for a heuristic search in the case that there are correlations among the samples of signal. We compare the performance of the new algorithm with various algorithms for uncorrelated and correlated sparsity. The results indicate the efficiency of the proposed methods.

*Keywords- Smoothed $l_0$; $l_1$ minimization; compressed sensing; reconstruction algorithm; correlated sparsity*

## I. INTRODUCTION

Sampling with some priori knowledge about desired signal has the benefit of sub-Nyquist rate. A well known priori knowledge is sparsity. A signal $s \in R^N$ is sparse in the domain $\Psi \in C^{N \times N}$, if it can be represented by a linear combination of a few atoms of the dictionary $\Psi$, i.e.

$$s = \Psi x = \sum_{i=1}^{N} \psi_i x_i \qquad (1)$$

where $\psi_i$ is the $i$-th column/atom of $\Psi$ and the vector $x \in R^N$, $x = [x_1, ..., x_N]^T$, is the $k$-sparse representation of $s$, if the $N$-element vector $x$ consists of at most $k$ non-zero elements ($k \ll N$) and $N - k$ zero elements. According to the compressed sensing theory, such signals can be sampled and compressed simultaneously [1]. Sampling is accomplished by linear non-adaptive random measurements from the whole sparse signal, which is expressed by

$$y = \Phi s = Ax \quad ; \quad A \triangleq \Phi \Psi \qquad (2)$$

where the elements of $y \in R^M$ are the compressively sensed samples, $\Phi \in R^{M \times N}$ is the measurement matrix ($M < N$), and $A \triangleq \Phi \Psi$ is defined as the recovery matrix. In order to reconstruct the Nyquist samples of the sparse signal $x$ from its compressed samples $y$ and the measurement matrix $\Phi$, the sparsest vector $x$ which satisfies (2) is desirable. It is computed by [1]:

$$\hat{x} = \arg\min_{x} \|x\|_0 \quad \text{subject to} \quad y = Ax \qquad (3)$$

where $\|x\|_0$ denotes the pseudo $l_0$ norm which is equal to the number of non-zero elements of $x$. In order to obtain the sparsest solution from (3), it is necessary to solve a combinatorial problem that is not feasible for the conventional applications. Alternatively, various algorithms have been proposed which have the same solutions as (3) under predefined conditions. These algorithms can be divided into three main categories: *a)* greedy and threshold based algorithms such as the matching pursuit (MP), CoSaMP, and IHT [2], [3], *b)* optimization based algorithms like the basis pursuit (BP), $l_1$ magic, and SL0 [4], [5], *c)* Bayesian framework based algorithms such as Bayesian compressed sensing (BCS) and BCS with Laplacian prior knowledge [6], [7].

An impressive alternative to $l_0$ minimization is obtained by $l_1$ relaxation, known as BP. It is given as

$$\hat{x} = \arg\min_{x} \|x\|_1 \quad \text{subject to} \quad y = Ax \qquad (4)$$

where $\|x\|_p = (\sum_{i=1}^{N} |x_i|^p)^{1/p}$ ($p \geq 1$), hence $\|x\|_1 = \sum_{i=1}^{N} |x_i|$.

This problem can be solved by linear programming and its solution will be the same as (3), if $\Phi$ is incoherent with $\Psi$ and also satisfies the Restricted Isometry Property (RIP). There are several random matrices that hold these conditions, if the number of their rows is of order $M = O(k \log(N/k))$ [8], [9]. In the case of noisy measurements, i.e. $y = \Phi s + n$, (4) has to be replaced with the Basis Pursuit De-Noising (BPDN) as

$$\hat{x} = \arg\min_{x} \|x\|_1 \quad \text{subject to} \quad \|y - Ax\|_2^2 \leq \varepsilon \qquad (5)$$

where $\varepsilon$ is the maximum energy of the measurement noise. For each value of $\varepsilon$, there exists a parameter $\lambda$ (regularization parameter) that makes the above equation equivalent to the following unconstrained optimization problem (known as Lasso problem):

$$\hat{x} = \arg\min_{x} \| y - Ax \|_2^2 + \lambda \| x \|_1 \qquad (6)$$

Another optimization based algorithm is smoothed $l_0$ (SL0). Concisely, it maximizes a Gaussian function of $x$ that gradually approaches the number of zero samples of $x$, when the variance of the Gaussian function ($\sigma^2$) smoothly nears zero [4]. That is, the SL0 solves the following problem for each value of $\sigma^2$:

$$\hat{x} = \arg\max_{x} f_\sigma(x) \quad \text{subject to} \quad y = Ax$$

where $f_\sigma(x_i) = e^{-\frac{x_i^2}{2\sigma^2}}$, thus $\| x \|_0 = N - \sum_{i=1}^{N} \lim_{\sigma \to 0} f_\sigma(x_i)$.

Furthermore, there are several algorithms developed for reconstruction of structured sparse signals with minimum number of measurements [10], [11]. The authors in [10] exploited the structure of wavelet coefficients for smooth piecewise signals in the reconstruction algorithm. Moreover, the basic algorithm has been generalized for reconstruction of block sparse signals in [10]. The multiband or equivalently smooth signal is an example of the signals which exhibit block sparsity in the Fourier domain. In [11], block sparse signals have been divided into equal length blocks to minimize the sum of $l_2$ norm ($l_1$-$l_2$ norm) of the blocks (instead of $l_1$ minimization in BP). Nevertheless, the authors in [10] and [11] used greedy based algorithms for generalization. The main disadvantage of these algorithms is the need for an oracle to predict the number of non-zero elements of the unknown vector. Moreover, a generalized version of the SL0 was presented in [12] which maximizes the $l_2$-$l_0$ norm of signal for block sparse recovery.

In this paper, we represent each element of a sparse vector $x$ as a function of two variables and update these variables instead of direct updating of $x$. The new method is inspired by the topology of radial basis function (RBF) networks and the annealing strategy used in the SL0 algorithm. We formulate the proposed algorithm, evaluate its performance, and compare it with several algorithms. We also extend the proposed model to exploit the correlation structure among the non-zero samples. Our generalized algorithm deploys the correlation matrix of the desired sparse signal to weight the updates of the extended variables. The results indicate that the proposed algorithm outperforms several existing algorithms in both cases of uncorrelated and correlated sparsity for the noiseless measurement, and has reasonable stability in the noisy measurements.

The rest of the paper is organized as follows. In Section II, we explain the proposed method and formulate the recovery algorithm by gradient descent method for uncorrelated signals. In Section III, the extended case of the proposed algorithm for reconstruction of correlated sparse signals is presented. In Section IV, we compare the performance of the new algorithm with other methods. Finally, Section V presents the conclusion and future works.

## II. PROPOSED ALGORITHM

Here, we consider uncorrelated sparse signals in which the correlation among the samples of the signal is zero. In the proposed method, each element ($x_i$) of an $N$-element vector $x = [x_1,...,x_i,...,x_N]^T$ is represented as a function of two main variables, $x_{vi}$ and $x_{di}$, named as "value" and "detector", respectively, and also a decreasing parameter $\sigma^2$:

$$x_i = x_{vi} e^{-\frac{(x_{di}-1)^2}{2\sigma^2}} \qquad i=1,...,N$$

By defining the vectors $x_v = [x_{v1},...,x_{vi},...,x_{vN}]^T$ and $x_d = [x_{d1},...,x_{di},...,x_{dN}]^T$, the sparse vector $x$ can be written as

$$x(x_v, \gamma) = \text{diag}(x_v)\gamma \qquad (7)$$

where $\text{diag}(x_v)$ is an $N \times N$ diagonal matrix with $x_{vi}$s as the diagonal elements, and $\gamma$ is an $N \times 1$ vector represented as

$$\gamma = [\gamma_1,...,\gamma_i,...,\gamma_N]^T = [e^{-\frac{(x_{d1}-1)^2}{2\sigma^2}},...,e^{-\frac{(x_{di}-1)^2}{2\sigma^2}},...,e^{-\frac{(x_{dN}-1)^2}{2\sigma^2}}]^T$$

where $0 \le \gamma_i \le 1$, $i=1,...,N$.

The $l_1$ norm of the proposed formulation for $x$ will be as

$$\| x \|_1 = \sum_{i=1}^{N} | x_{vi} | \gamma_i \qquad (8)$$

At first glance, the above formulation for the $l_1$ norm reminds of the reweighted $l_1$ algorithm [13], that assigns weights $w \in R^N$ to the elements of the vector $x$ in each iteration in a way that, for high values of $x_i$, the weight is small and for small values of $x_i$, the weight is large (i.e. $w \propto 1/|x_i|$). This iterative reweighting makes the behavior of $l_1$ norm more similar to the discontinuous $l_0$ function. However, unlike the reweighted $l_1$ algorithm, in our algorithm the weights ($\gamma_i$ s) are in the interval [0,1] and large weights (maximum 1) are assigned to non-zero elements whereas small weights (minimum 0) are assigned to the zero elements. Although, this kind of weighting is not desirable in the $l_1$ minimization (we will eliminate its effect as shown later), but it predicts the locations of non-zero elements and it can be a useful operator to enhance the selection property of the Lasso. Especially, the presence of these "oracle variables" ($\gamma$) is crucial for the extended algorithm in the correlated case in which the non-diagonal elements of $\text{diag}(x_v)$ also participate in the optimization process (will be explained in Section III).

Table I summarizes the steps of the proposed algorithm. Similar to SL0, our algorithm starts from a sufficiently large value of $\sigma^2$ ($\sigma^2_{\max}$), in which $\gamma_i \simeq 1$ for all elements of the vector $x_v$. In a decreasing procedure for $\sigma^2$, for each $\sigma^2$, we update $x_v$ and $x_d$ by the gradient descent method, until the distance of the estimated measure $A\hat{x}$ from the exact measurement $y$ becomes less than $\beta\sigma^2$, where $\beta$ is an appropriate constant. $\beta\sigma^2$ is obtained empirically. Indeed, finding a convergence guarantee of the aforementioned stop condition is difficult, specially at low values of $\sigma^2$. Hence, we use an auxiliary condition so that if the number of iterations in the inner loop (*i*) is larger than the predefined value $I_{max}$, then the algorithm exits the inner loop and decreases $\sigma^2$. After one of the convergence conditions, $\|A\hat{x}-y\|_2 < \beta\sigma^2$ or $i > I_{max}$, is satisfied for a specific $\sigma^2$ (inner loop of Table I), $\sigma^2$ is updated by a constant step size $\tau$ as $\sigma^2 = \sigma^2 - \tau$, until $\sigma^2 = \sigma^2_{min}$ where $\sigma^2_{min}$ is a predefined value.

During the recovery process, the elements of $x_v$ approach the exact values of $x$ and $\gamma_i$ s gradually decrease from 1 to 0 for zero samples of the sparse signal to downplay the effect of the corresponding columns in the matrix diag($x_v$), i.e. zero samples, in the square error minimization (atom pruning). The other values of $\gamma_i$ s try to remain 1 for non-zero samples of *x*. At the end of algorithm, these binary tight values for $\gamma_i$ s are desirable. However, in practice, this ideal case is not possible and $\gamma$ makes some fuzzy decisions about the non-zero elements and even its final values for zero samples are not absolutely zero. Nevertheless, an appropriate update for $x_d$ can help in approaching the desired values for $\gamma$.

Here, we use gradient descent method for solving the optimization problem (6) considering the proposed decomposition (7). Updating of $x_v$ is performed based on both the least square error $\|Ax-y\|_2$ and the regularization term $\|x\|_1$ of the Lasso. These functions are convex with respect to $x_v$. On the other hand, $x_d$ is updated only based on the $l_1$ norm of *x*, that is concave with respect to $x_d$ in the interval $[1-\sigma, 1+\sigma]$. The gradients of $\|x\|_1$ and $\|Ax-y\|_2$ are derived as:

$$\delta_d = \frac{d\|x\|_1}{dx_d} = -\frac{(x_d-1)}{\sigma^2}.|x_v|.\gamma \qquad (9)$$

$$\delta_{v_1} = \frac{d\|x\|_1}{dx_v} = \text{sign}(x_v).\gamma \qquad (10)$$

$$\delta_{v_2} = \frac{d\|y-Ax\|_2^2}{dx_v} = -2A^T(y-Ax).\gamma \qquad (11)$$

where $\delta_d$, $\delta_{v_1}$, and $\delta_{v_2}$ are the *N*-dimensional gradient vectors and *U.V* indicates element by element multiplication of vectors *U* and *V*.

*Remarks:*

- Since $\sigma^2$ appears in the denominator of (9), we select the corresponding step size of $\delta_d$ ($\mu_d$ in Table I) as a factor of $\sigma^2$ to prevent the increase of $\delta_d$ in low values of $\sigma^2$.

- As mentioned before, since the Gaussian function $\gamma_i$ prevents from zero enforcing property of sign($x_v$) (see [14] for an elaboration) in (10), we choose the step size of $\delta_{v_1}$ as $\mu_{v_1} = \mu_0./\gamma$ to compensate for this unwanted effect. $\mu_0$ is a predefined parameter.

TABLE I. PROPOSED ALGORITHM

**inputs**: $y, A = \Phi\Psi$   **output**: $x = \hat{x}$

**initialization**:
$\quad x_v^{(0)} \leftarrow \mathbf{0}_{N\times 1}$, $x_d^{(0)} \leftarrow \mathbf{-1}_{N\times 1}$
$\quad \mu_{v_1}, \mu_{v_2}, \mu_d, \tau$ (step sizes)
$\quad \sigma^2_{min}, \sigma^2_{max}, I_{max}$

**while** $\sigma^2 > \sigma^2_{min}$
$\quad i = 0$
$\quad$ **while** $\|A\hat{x}-y\|_2 > \beta\sigma^2$ **and** $i < I_{max}$

$\quad\quad x_v^{(n+1)} = x_v^{(n)} - \mu_{v_2}\delta_{v_2} - \mu_{v_1}\delta_{v_1}$

$\quad\quad x_d^{(n+1)} = x_d^{(n)} + \mu_d \delta_d$

$\quad\quad \hat{x} = x_v^{(n+1)}\gamma \qquad (\gamma \triangleq e^{-\frac{(x_d-1)^2}{2\sigma^2}})$

$\quad\quad i = i+1$
$\quad$ **end**
$\quad \sigma^2 = \sigma^2 - \tau$
**end**

## III. EXTENSION TO CORRELATED SPARSITY

Here, our goal is to recover the correlated non-zero elements of the sparse vector with minimum number of measurements using the correlation matrix (*C*) in the reconstruction algorithm. In this case, we replace diag($x_v$) with the matrix $X_v \in R^{N\times N}$ and represent the elements of *x* by linear equations as follows ($x = X_v \gamma$):

$$\begin{pmatrix} x_1 \\ x_2 \\ \vdots \\ x_N \end{pmatrix}_{N\times 1} = \begin{pmatrix} x_{v_{11}} & x_{v_{12}} & \cdots & x_{v_{1N}} \\ x_{v_{21}} & x_{v_{22}} & \cdots & x_{v_{2N}} \\ \vdots & & \ddots & \vdots \\ x_{v_{N1}} & \cdots & & x_{v_{NN}} \end{pmatrix}_{N\times N} \begin{pmatrix} e^{-\frac{(x_{d1}-1)^2}{2\sigma^2}} \\ \vdots \\ e^{-\frac{(x_{dN}-1)^2}{2\sigma^2}} \end{pmatrix}_{N\times 1} \qquad (12)$$

Similar to the uncorrelated case, $X_v$ minimizes both of the square error $\|Ax-y\|_2$ and the $l_1$ norm $\|x\|_1$ and $x_d$ only participates in sparsifing the solution. It can be shown that the gradients are derived as

$$\delta_d = \frac{d\|x\|_1}{dx_d} = -X_v^T \text{sign}(x).\frac{(x_d-1)}{\sigma^2}.\gamma \qquad (13)$$

$$\delta_{v_1} = \frac{d\|x\|_1}{dX_v} = \text{sign}(x)\gamma^T \qquad (14)$$

$$\delta_{v_2} = \frac{d\|y-Ax\|_2^2}{dX_v} = -A^T(y-Ax)\gamma^T \qquad (15)$$

where $\delta_{v_2}$ and $\delta_{v_1}$ are $N\times N$ gradient matrices and $\delta_d$ is an $N\times 1$ vector.

The increase in the number of optimization variables (from $2N$ in the uncorrelated case to $N^2+N$ in the correlated case) leads to an extended search area which can convert the original algorithm to an intractable one. For the sake of tractability, we weight the non-diagonal elements of $X_v$ by the correlation matrix $C$ in the updating process of the least square error, i.e.

$$X_v^{(n+1)} = X_v^{(n)} - \mu_{v_2} C \cdot \delta_{v_2} - \mu_{v_1} \delta_{v_1} \quad (16)$$

As an interpretation of this weighting by $C$, high correlation values in the matrix $C$ encourage the corresponding $x_{v_{ij}}$ s to participate in the squared error minimization and low values in the correlation matrix $C$ discourage the corresponding entries of $X_v$. Hence, a numerous number of the non-diagonal variables never participate in the optimization process, because the correlation matrix of the sparse signals is sparse too. In other words, when $x_{di}$ detects the $i$-th sample of $x$ as a non-zero element, equivalently it selects the $i$-th column of $X_v$ and allows this sample to affect the other non-zero samples by adding $x_{v_{ij}}$ s, $j=1,...,N, j \neq i$, which are updated depending on their correlations with $x_{v_{ii}}$. Obviously, when $C=I$, the above rules will be same as the uncorrelated case.

*Remarks:*

- Since $\gamma^T$ do not produce appropriate weights for sign($x$) (Section II), hence we replace $\gamma^T$ with $[(1./\gamma)-1]^T$ in (14) and use the recommended step size $\mu_d$ in the uncorrelated case.

- In case of the sparsity in the other domains, like Fourier or wavelet domains, we need the correlation matrix of the Fourier/wavelet coefficients that can be computed by $C_G = GCG^T$, where $G$ is the corresponding matrix of Fourier, wavelet, or any another linear transformation [15].

## IV. EXPERIMENTS AND RESULTS

In this section, we evaluate the performance of the proposed algorithm and compare it with other methods[1] in both cases of uncorrelated and correlated sparsity.

### A. Uncorrelated Sparsity

In this experiment, we used a typical random sparsity, in which the locations of non-zero elements of the sparse vector were selected randomly with equal probability and the values of non-zero elements were chosen from i.i.d. Gaussian distributions. In order to measure this signal, we utilized Gaussian random matrices with the normalized columns that satisfy the RIP condition with high probability in the given sparsity level. In both experiments (uncorrelated and correlated sparsity), we used the distortion metric $\|x-\hat{x}\|_2^2 / \|x\|_2^2$ to evaluate the performance of different recovery algorithms.

---
[1] Codes are available in: http://sites.google.com/site/igorcarron2/cscodes

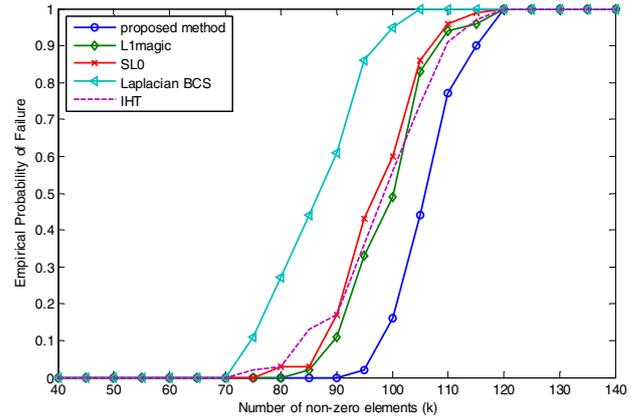

Figure 1. Probability of failure versus various levels of sparsity, $N=500$, and $M=250$.

*1) Phase transition:* Here, we calculated the distortion of several algorithms in the recovery of 100 different randomly generated sparse signals for a given range of sparsity. The random measurement matrices were different in each trial. The recovery was considered as failed, if the distortion values are larger than $10^{-3}$, otherwise as successful. The probability of failure is obtained as the ratio of the number of failures in 100 trials. The parameters of this experiment were selected as

$\mu_{v_1} = \sigma^2./(10^4 \gamma)$, $\mu_{v_2} = 0.01$, $\mu_d = \sigma^2/10^4$, $\beta=0.1$, $\tau=0.1$, $I_{max}=2000$, $\sigma_{min}^2=1$, $\sigma_{max}^2=10$

Fig. 1 depicts the failure probability of different methods in the recovery of sparse signal with the length $N=500$ from $M=250$ noiseless measurements for various number of non-zero elements ($k$). The results indicate considerable improvement of the proposed method in comparison with $l_1$-magic [5], SL0 [4], Laplacian BCS [7], and IHT [3] algorithms. We observe that the new method bears the high levels of sparsity for successful recovery with overwhelming probability.

*2) Stability:* We repeated the previous experiment in the noisy case in which the measurement is contaminated by white Gaussian noise. Fig. 2 shows the mean of recovery distortion for various algorithms versus signal to noise ratio (SNR) for two different levels of sparsity. By comparison of different methods, the following two important points can be resulted:

a) The IHT algorithm as a greedy method has different stabilities depending on the sparsity level. In this experiment, for low levels, it gives the best performance, while it has great degradation in high levels.

b) The proposed algorithm has better stability compared to SL0 and $l$1-magic as the optimization based algorithms. Also, it is more stable than the BCS with Laplacian prior at high SNRs in which the distortion is reasonable.

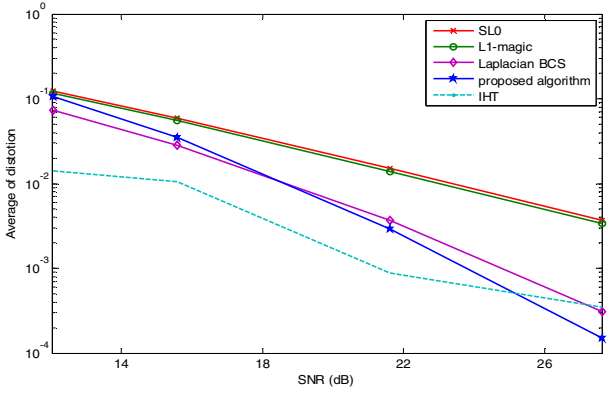

2-a) Low level of sparsity ($k = 15$)

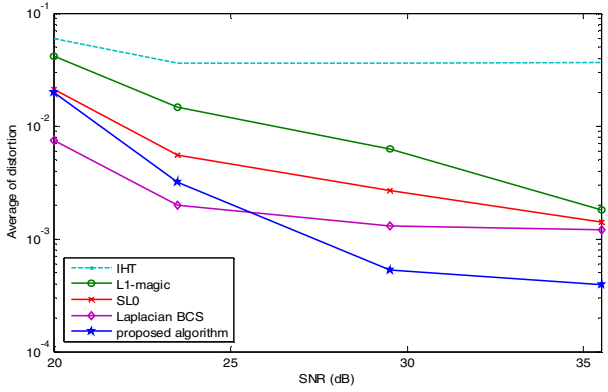

2-b) High level of sparsity ($k = 90$)

Figure 2. Stability of the various algorithms against noise at two different levels of sparsity.

### B. Correlated Sparsity

We use the following exponential correlation matrix (17) to form an approximation of the correlation matrix of the desired sparse vector. It is obvious that the extended algorithm will achieve better performance, if the exact information is available. Here, the experiments are performed in the two sparsity domains of time and wavelet.

$$C = \begin{pmatrix} 1 & \alpha & \alpha^2 & \cdots & \alpha^{N-1} \\ \alpha & 1 & \alpha & \cdots & \alpha^{N-2} \\ \vdots & & \ddots & & \vdots \\ & & & & \alpha \\ \alpha^{N-1} & \cdots & & \alpha & 1 \end{pmatrix}_{N \times N} ; 0.5 < \alpha < 1 \quad (17)$$

*1) Time domain (correlated block sparsity):* In this experiment, we evaluate the performance of the proposed method for the correlated block sparse signals. Here, a $500 \times 1$ vector is divided into 50 blocks; each block contains $L = 10$ samples and non-zero blocks have correlation among their samples. The samples of non-zero blocks can be generated by $x_b = Q\sqrt{\Lambda} n$ where $n$ is an uncorrelated $10 \times 1$ vector (generated from white Gaussian noise in this experiment). $Q$

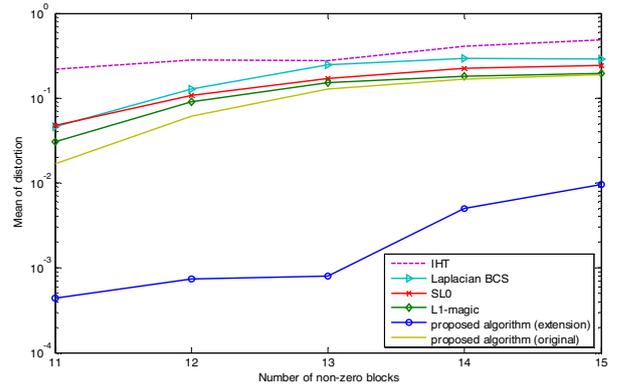

Figure 3. Mean of distortion in recovery of correlated block sparsity.

and $\Lambda$ are obtained from the decomposition of the correlation matrix $C = Q\Lambda Q^T$ where $Q$ is the matrix containing eigen vectors and $\Lambda$ is a diagonal matrix with eigen values as its elements. Hence, $x_b$ will be a correlated $10 \times 1$ vector with the covariance matrix $Q\Lambda Q^T$ [15]. We selected $C$ as (17) with $\alpha = 0.99$ for each non-zero block and used the following matrix as the overall correlation matrix of the desired signal:

$$C_{total} = \begin{pmatrix} C_{L \times L} & 0_{L \times L} & \cdots & 0_{L \times L} \\ 0_{L \times L} & C_{L \times L} & & 0_{L \times L} \\ \vdots & & \ddots & \vdots \\ 0_{L \times L} & \cdots & 0_{L \times L} & C_{L \times L} \end{pmatrix}_{N \times N}$$

where $0_{L \times L}$ is the matrix of zeros and $C_{total}$ is $500 \times 500$ matrix with $10 \times 10$ matrices ($C$) on the main diagonal. We depicted the recovery distortion averaged on 100 trials for different number of non-zero blocks. The number of measurements was $M = 250$ and the parameters were set to:

$\mu_{v_1} = 10^{-5}$, $\mu_{v_2} = 0.01$, $\mu_d = \sigma^2/10^2$, $\sigma_{\min}^2 = 0.9$, $\sigma_{\max}^2 = 50$,
$I_{\max} = 2000$, $\beta = 0.1$, $\tau = 0.1$

The results shown in Fig. 3 indicate the significant performance improvement of the proposed method in the high sparsity levels (11 to 15 non-zero blocks corresponding to 110 to 150 non-zero samples) in comparison with the IHT [3], SL0 [4], $l_1$ magic, and Laplacian BCS [7] methods.

*2) Wavelet domain:* We examined the extended algorithm in the wavelet domain. Piecewise smooth signals have a compressible representation in the wavelet domain. We used HeaviSine signal as a one dimensional piecewise smooth signal. Sampling was accomplished by a Gaussian random matrix. We applied third stage Haar wavelet transform [16] as the atoms of dictionary $\Psi$ at the recovery stage. Using higher levels of wavelet transform results in more sparse representation because of energy compaction in the base scale, while the dynamic range of the coefficients increases which renders the recovery process challenging. An exponential correlation matrix like (17) with $\alpha = 0.99$ was assumed for the HeaviSine and the correlation among its coefficient in wavelet domain was computed by the mentioned relation in Section III. The length of the signal was $N = 1024$ and the number of measurements was $M = 330$ and noiseless.

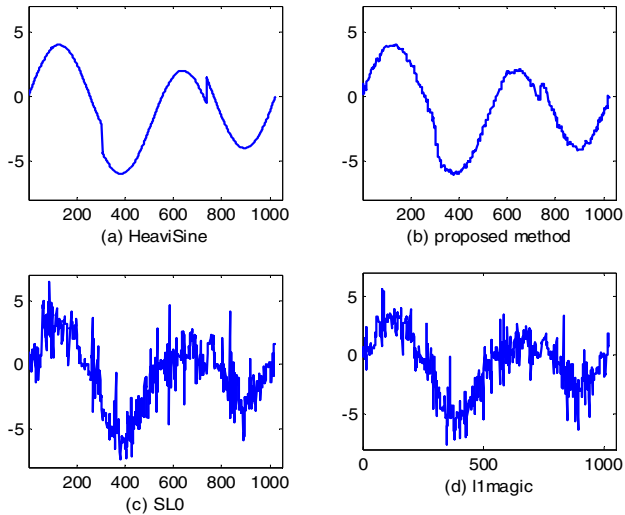

Figure 4. Reconstruction of HeaviSine signal with different algorithms ($N$=1024, $M$=330).

Fig. 4 shows the original signal and the reconstructed signals using the proposed and two other basic algorithms. The reconstruction distortion was $4.2 \times 10^{-3}$ for the proposed method, 0.19 for SL0 [4], and 0.16 in $l_1$-magic methods [5]. These results illustrate the profound effect of the structural reconstruction.

## V. CONCLUSION AND FUTURE WORK

We presented a new reconstruction algorithm for sparse signals by defining its samples as functions of two variables. The first variable was used for solving the same problem as the Lasso, while the second variable played the role of an oracle for the first variable. Moreover, we extended the original algorithm to a heuristic recovery method by using the correlation matrix as additional priori knowledge about the samples. We experimentally showed that the increase in the number of variables in the optimization problem, in both proposed method, renders high quality reconstructions respect to the basic algorithms of the different classes. In the future work, we will search for providing a guide on the optimal parameters selection, explore the optimization methods to achieve faster versions, especially in the correlated case, and compare the results with the specialized algorithms for structural recovery.

ACKNOWLEDGMENT

The authors are grateful to Dr. Gholamhossein Gholami for his insightful discussions.

REFERENCES

[1] R. G. Baraniuk, "Compressive sensing", *IEEE Signal Processing Magazine,* vol. 24, no. 4, pp. 118–121, Jul. 2007.

[2] D. Needell and J. Tropp "CoSaMP: Iterative signal recovery from incomplete and inaccurate samples", *Elsevier, Applied and Computational Harmonic Analysis*, vol. 26, no. 3, pp. 301-321, May 2009.

[3] T. Blumensath and M. Davies, "Iterative hard thresholding for compressed sensing", *Elsevier, Applied and Computational Harmonic Analysis*, vol. 27, no. 3, pp. 265-274, Nov. 2009.

[4] H. Mohimani, M. Babaie-Zadeh, and C. Jutten, "A fast approach for overcomplete sparse decomposition based on smoothed ℓ0 norm", *IEEE Trans. Signal. Proc.*, vol. 57, no. 1, pp. 289 - 301, Jan. 2009.

[5] E. Candes and J. Romberg, "ℓ1-magic: Recovery of sparse signals via convex programming," URL:www.acm.caltech.edu/l1magic/downloads/l1magic.pf, 2005.

[6] S. Ji, Y. Xue, and L. Carin, "Bayesian compressive sensing," *IEEE Trans. Signal Process.*, vol. 56, no. 6, pp. 2346–2356, Jun. 2008.

[7] S. Babacan, R. Molina, and A. Katsaggelos, "Bayesian compressive sensing using Laplace priors", *IEEE Tran. Image Proc.*, vol. 19, no. 1, pp. 1057-7149, Jan. 2010.

[8] D. L. Donoho, "Compressed sensing", *IEEE Trans. Info. Theory*, vol. 52, no. 4, pp. 1289-1306, Apr. 2006.

[9] E. J. Candes, J. K. Romberg, and T. Tao, "Robust uncertainty principles: exact signal representation from highly incomplete frequency information", *IEEE Trans. Info. Theory*, vol. 52, no. 2, pp. 489–509, Feb. 2006.

[10] R. Baraniuk, V. Cevher, M. Duarte, and C. Hegde, "Model-based compressive sensing", *IEEE Trans. Info. Theory*, vol. 56, no. 4, pp. 1982-2001, Apr. 2010.

[11] M. Stojnic, F. Parvarersh, and B. Hassibi, "On the reconstruction of block-sparse signals with an optional number of measurements", *IEEE Trans. Signal. Proc.*, vol. 57, no. 8, pp. 3075 - 3085, Aug. 2009.

[12] S. Hamidi, M. Babaie-Zadeh, and C. Jutten, "Fast block-sparse decomposition based on SL0", Springer, *Lecture Notes in Computer Science*, vol. 6365, pp. 426-433, 2010.

[13] E. Candes, M. Wakin, and S. Boyd, "Enhancing sparsity by reweighted L1 minimization", Springer, *Journal of Fourier Analysis and Applications*, vol. 14, no. 5-6, pp.877-905, 2008.

[14] Y. Chen, Y. Gu, and A. Hero, "Sparse LMS for system identification", *Proc. IEEE ICASSP*, 2009.

[15] A. Papoulis, and S. U. Pillai, *Probability, Random Variables, and Stochastic Processes*. New York: McGraw-Hill, 2002.

[16] M. Weeks, *Digital Signal Processing Using Matlab and Wavelets*. Infinity Science Press, 2007.